\documentclass[fleqn,10pt]{wlscirep}
\usepackage[utf8]{inputenc}
\usepackage[T1]{fontenc}
\usepackage{caption}
\usepackage{subcaption}
\usepackage{amsmath}
\usepackage{nccmath}

\usepackage{mwe} 
\usepackage{lipsum} 

\title{Physics inspired compact modelling of BiFeO$_3$ based memristors for hardware security applications}

\author[1,*]{Sahitya Yarragolla}
\author[2, 3 +]{Nan Du}
\author[1]{Torben Hemke}
\author[2,3]{Xianyue Zhao}
\author[2,3]{Ziang Chen}
\author[4]{Ilia Polian}
\author[1,$\dagger$]{Thomas Mussenbrock}

\affil[1]{Chair of Applied Electrodynamics and Plasma Technology, Ruhr University Bochum, Germany}
\affil[2]{Institute for Solid State Physics, Friedrich Schiller University Jena (FSUJ), Germany}
\affil[3]{Department of Quantum Detection, Leibniz Institute of Photonic Technology (IPHT), Germany}
\affil[4]{Institute of Computer Science and Computer Engineering, University of Stuttgart, Stuttgart, Germany}
\affil[*]{sahitya.yarragolla@rub.de}

\affil[+]{nan.du@uni-jena.de}

\affil[$\dagger$]{thomas.mussenbrock@rub.de}

\begin{abstract}
With the advent of the Internet of Things, nanoelectronic devices or memristors have been the subject of significant interest for use as new hardware security primitives. Among the several available memristors, BiFe$\rm O_{3}$\,(BFO)-based electroforming-free memristors have attracted considerable attention due to their excellent properties, such as long retention time, self-rectification, intrinsic stochasticity, and fast switching. They have been actively investigated for use in physical unclonable function (PUF) key storage modules, artificial synapses in neural networks, nonvolatile resistive switches, and reconfigurable logic applications. In this work, we present a physics-inspired 1D compact model of a BFO memristor to understand its implementation for such applications (mainly PUFs) and perform circuit simulations. The resistive switching based on electric field-driven vacancy migration and intrinsic stochastic behaviour of the BFO memristor are modelled using the cloud-in-a-cell scheme. The experimental current-voltage characteristics of the BFO memristor are successfully reproduced. The response of the BFO memristor to changes in electrical properties, environmental properties (such as temperature) and stress are analyzed and consistent with experimental results. 
\end{abstract}
\begin{document}

\flushbottom
\maketitle
%
%
\thispagestyle{empty}

\section*{Introduction}

It is highly appreciated how the Internet of Things (IoT) has inevitably integrated into our lives, making it more convenient and efficient. However, with the expansion and vast diffusion of connected devices in the IoT, cybersecurity concerns have also increased. The privacy of individuals, companies and institutions have been highly compromised \cite{Chen2019,Weber2010}. Unfortunately, the classical security solutions (software-level mathematical or algorithmic solutions) are no longer sufficient to secure modern-day applications. The increasing physical and side-channel attacks necessitate the need for alternative solutions \cite{Rajendran2021}.

Researchers and engineers have shifted their focus towards finding hardware-level solutions to address security-related challenges in recent times. The hardware-level solutions include the new nano-electronic devices, such as memristive devices or memristors, spintronics, or carbon nanotubes\cite{Rose2014}. Explicitly, memristive devices are foreseen as promising candidates for future hardware security applications mainly because of their special properties, such as low power consumption, scalability to the nano grade, fast switching, large off/on ratio, good endurance and reliability \cite{Yachuan2019,Du2021,Gokulnath2021}. Also known as the resistive switching random access memory (ReRAMs), these memristive devices are two-terminal devices whose resistance can be changed by applying a suitable electrical input. Apart from the features mentioned above, the switching mechanisms in these devices are intrinsically stochastic, which make these devices highly suitable for hardware security applications like physical unclonable functions (PUFs)\cite{Mesbah2019,Ibrahim2022}, true random number generators (TRNGs)\cite{Jiang2017}, and hash functions\cite{Azriel2017}. 

So far, many devices have been reported that exhibit resistive switching behaviour; however, in the present work, we focus on the devices where the resistive switching is triggered by ionic motion driven by an electric field\cite{Waser2007}. These devices can be either filamentary-type devices involving filaments' formation or interface-type (also called non-filamentary) devices involving the movement of charged defects. As mentioned by Du \textit{et al.}\cite{Du2019electroformingfree,Du2021}, the high currents induced in filamentary devices during the electroforming process can damage or destroy the device via thermodynamic dielectric breakdown, reducing the reliability of the device. In order to avoid the electroforming process, interface-type devices such as $\rm BiFeO_{3}$ (BFO)-based memristors\cite{Shuai2013_key,Du2018, You2014Adv}, double-barrier memristive devices (DBMD)\cite{Hansen2015} are preferred. The BFO memristive devices have been intensively studied in the memristive community because they exhibit excellent characteristics such as electroforming free switching, long retention time, good endurance, and also offer multistage switching. These features make the BFO device highly recommended for implementing future hardware security applications such as PUFs and TRNGs.

Furthermore, the development of existing and new memristive devices for hardware security applications requires a precise understanding of their physical behaviour. It is often challenging to determine the exact switching mechanism using experimental or diagnostic methods. Therefore, simulation models are developed that can contribute significantly to understanding the behaviour of such devices. On the one hand, multi-dimensional computational models (such as 3D kinetic Monte-Carlo)  are exploited for an in-depth understanding of resistive switching and moderately include real-world devices' physical and chemical processes and stochastic behaviour~\cite{Dirkmann2016,Dirkmann2018,Abbaspour2016}. However, they are computationally very expensive and, therefore, cannot be used for performing circuit simulations. On the other hand, there are compact or concentrated models based on mathematical formulae. They are fast but do not include the physical and chemical processes in the device, and often, they do not include the intrinsic stochasticity found in these devices~\cite{Solan2017,Ambrogio2014,Bengel2020}.

\begin{figure}[!t]
                \centering             \includegraphics[width=0.95\textwidth]{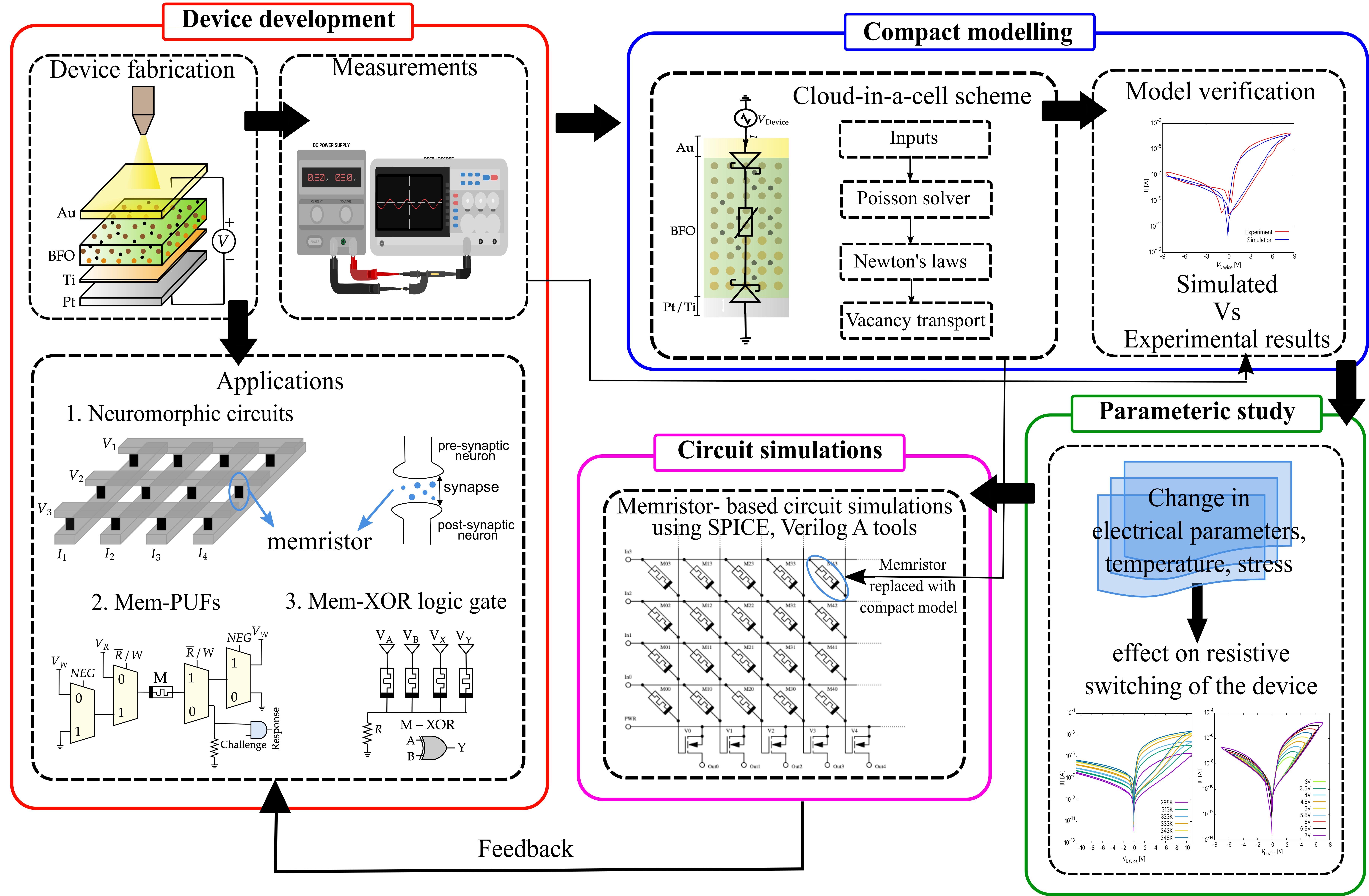}
                \caption{A flowchart illustrating the goal of the proposed work.}
                \label{fig:intro1}
            \end{figure}

Unlike the state-of-the-art models, we propose a circuit simulator-compatible distributed model for BFO memristor that considers the advantages of both the models mentioned above. It is a one-dimensional (1D) compact model including more or less realistic physics and the experimentally observed stochastic behaviour i.e., the cycle-to-cycle (C2C) variability and device-to-device (D2D) variability observed in BFO memristors. A kinetic Cloud-in-a-cell (CIC)\cite{Laux1996,Yarragolla2022} scheme is used to simulate the resistive switching mechanism based on the ion/vacancy transport. Although it is a distributed model, because we resolve it in a 1D space, it is computationally less demanding and fast. The model is primarily considered to provide an interface between circuit designers and device developers, as shown in Fig.\ref{fig:intro1}\cite{Saha2015}. It is used to explore the electrical properties of BFO as entropy sources and the effects of physical variables such as temperature and voltage on the entropy sources. The model can be extended further to investigate the performance of BFO memristive devices for hardware security applications by performing circuit simulations of memristive-PUFs or memristive-TRNGs with a SPICE-like circuit simulator.

The manuscript is divided into three sections.  First, the simulation approach is discussed in detail, explaining the BFO memristive device and its current mechanisms. Then, the simulation results based on the experimentally determined electrical parameters of BFO are discussed and compared with the experimental findings. Finally, an overall summary and significant findings from the current work are provided in the conclusion section.

\section*{Simulation approach}

    \begin{figure}[!t]
     \centering
     \begin{subfigure}[b]{0.4\textwidth}
         \centering
         \includegraphics[width=0.91\textwidth]{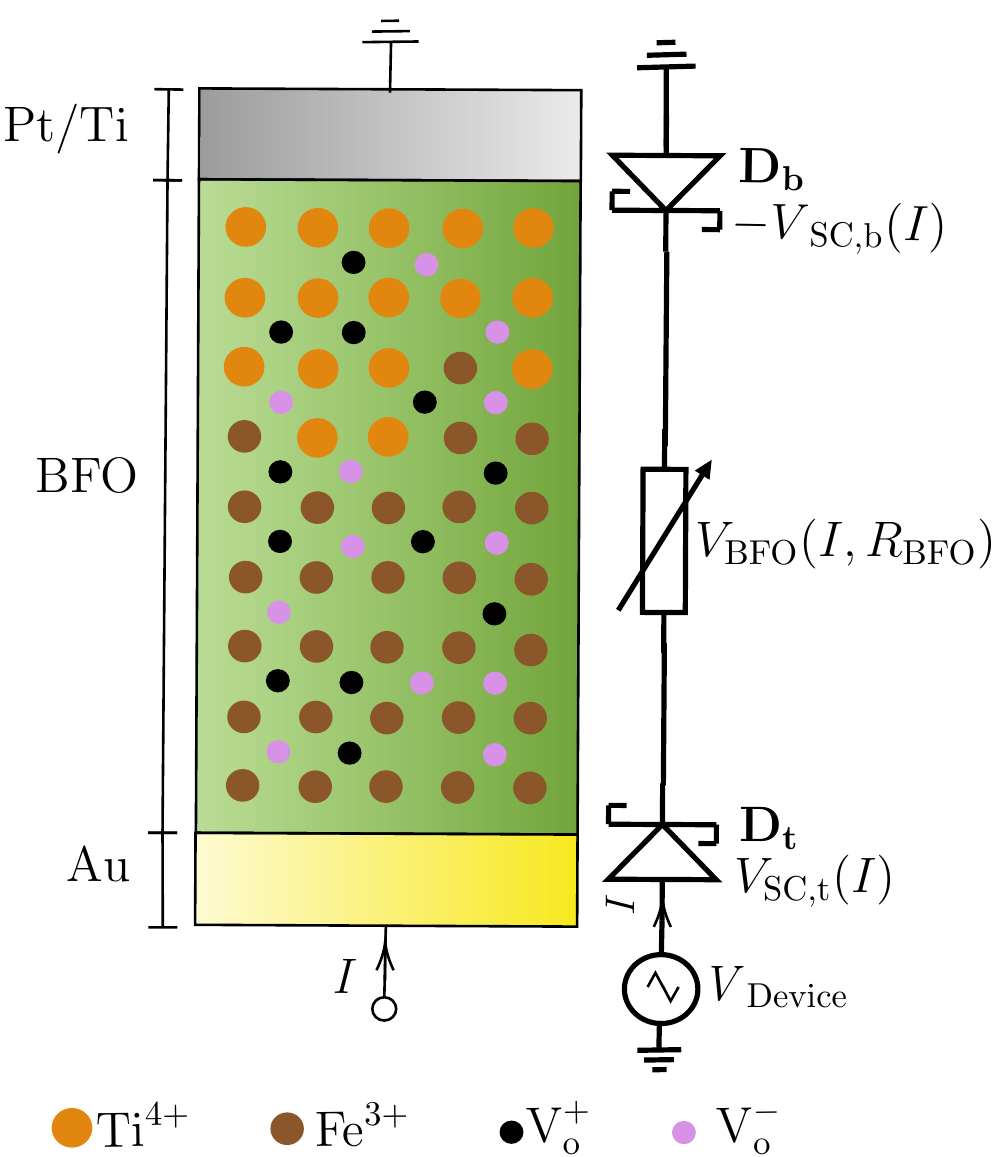}
         \caption{}
         \label{fig:Sim1a}
     \end{subfigure}
     \hspace{0.5cm}
     \begin{subfigure}[b]{0.46\textwidth}
         \centering      \includegraphics[width=\textwidth]{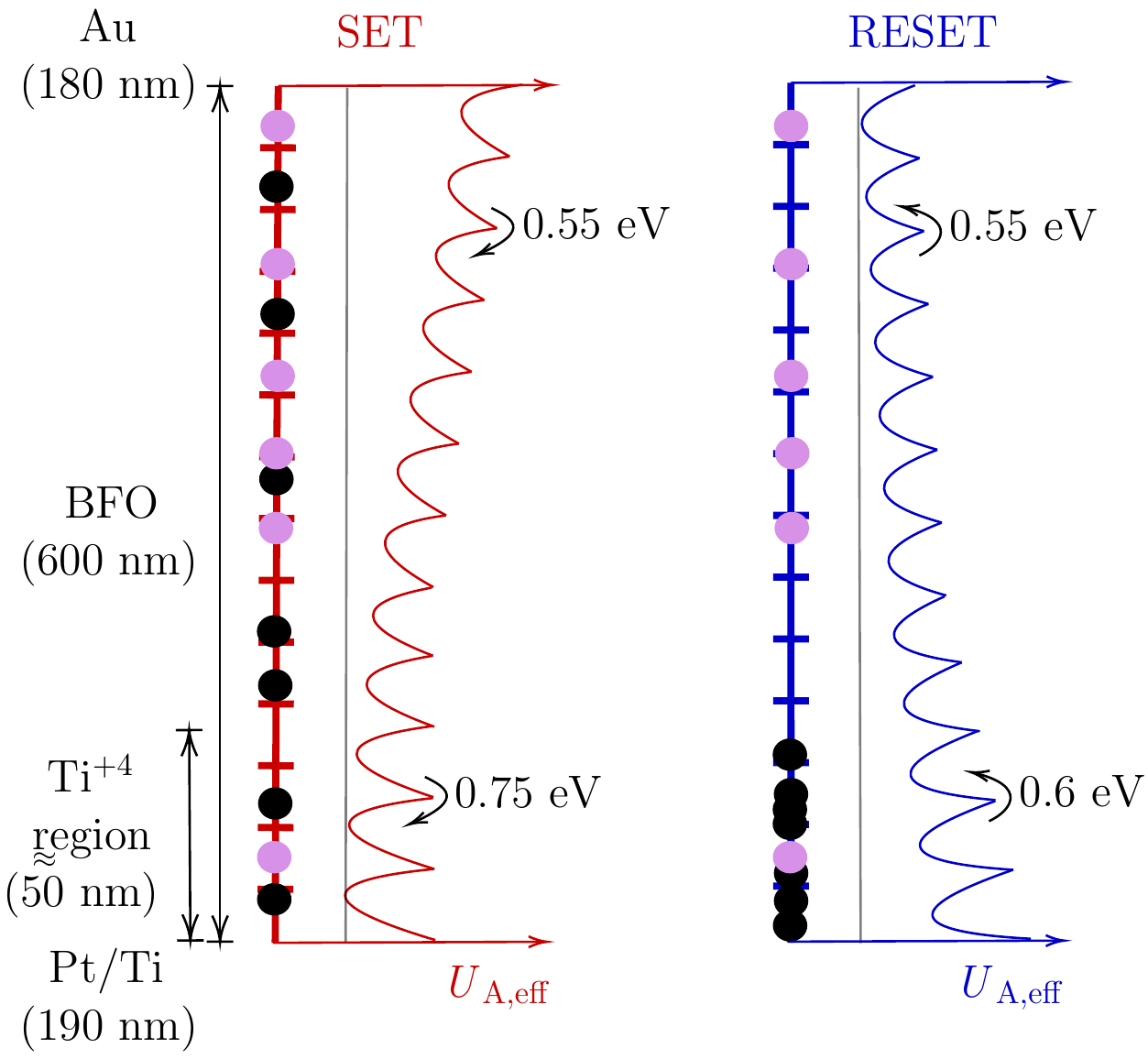}
         \caption{}
         \label{fig:Sim1b}
     \end{subfigure}
        \caption{Simulation scenario of BFO memristor. (a) a $\rm Au/BiFeO_{3}\,(BFO)/Pt/Ti$ memristive device and it's equivalent circuit with back-back schottky diodes and a variable resistor. (b) Ion transport and the potential structure across a 1D BFO memristor for set and reset process.}
        \label{fig:Sim1}
\end{figure}

A polycrystalline perovskite structured $\rm Au/BiFeO_{3}\,(BFO)/Pt/Ti$ memristive device and its equivalent circuit are shown in Fig.\ref{fig:Sim1a}\cite{Du2018}. A BFO memristive device is regarded as an interface-type memristive device with $\rm BiFeO_{3}$ as the primary layer, a rectifying Au/BFO top Schottky contact with nearly unflexible barrier height ($\rm D_{t}$), and a rectifying and/or Ohmic BFO/Pt/Ti bottom Schottky contact with flexible barrier height ($\rm D_{b}$). Previous studies have shown that the resistive switching behaviour in BFO can be described using the ferroelectric polarisation mechanism~\cite{Choi2009,Can2011,Lei2016} or the oxygen vacancy migration mechanism~\cite{You2014,Du2018}. However, as reported by You \textit{et al.}~\cite{You2014} and Du \textit{et al.}~\cite{Du2018}, the resistive switching behaviour in BFO is primarily due to the migration of positively charged defects, and the ferroelectric switching can be excluded based on polarisation–electric-field measurements. The positively charged defects are identified as the oxygen vacancies ($\rm V_{O}^{+}$) present at the interstitial sites. When a positive write biased is applied to the device, the $\rm V_{O}^{+}$ vacancies migrate to the bottom BFO/Pt interface under the influence of the electric field, thereby lowering the barrier height of $\rm D_{b}$ and setting the device to low resistance state (LRS). These $\rm V_{O}^{+}$ vacancies migrate back to their equilibrium positions for a negative write bias, re-setting the device to an high resistance state (HRS). The barrier height of $\rm D_{t}$ is almost unaffected by the $\rm V_{O}^{+}$ vacancies migration.

The other fixed ions in the BFO are the $\rm Fe^{3+}$ ions and $\rm Ti^{4+}$ ions. These two fixed ions do not participate directly in the switching process but mainly determine the activation energy ($U_{\rm A}$) in the BFO. The $\rm Ti^{4+}$ ions replace the $\rm Fe^{3+}$ ions close to the BFO/Pt interface. Due to this, the activation energy is not uniform but increases gradually from 0.55\,eV at Au/ BFO interface to 0.75\,eV at BFO/PT interface, depending on the occupation of $\rm Ti^{4+}$ ions. It is important to note here that with the diffusion of $\rm Ti^{4+}$ ions into the $\rm BiFeO_{3}$ layer, the activation energy near the bottom electrode is increased to the point where vacancies can be easily trapped, thus improving the retention and endurance of the BFO memristor\cite{You2015}.

According to You \textit{et al.}~\cite{You2014} and Du \textit{et al.}~\cite{Du2018}, the $\rm V_{O}^{+}$ migration can be explained as: (a) the back and forth movement of $\rm V_{O}^{+}$ vacancies in the presence of an electric field with a certain drift velocity and (b) the trapping or de-trapping of $\rm V_{O}^{+}$ vacancies in traps or potential wells formed by the $\rm Ti^{4+}$ fixed ions. A comparable $\rm V_{O}^{+}$ migration is implemented in this paper using the CIC scheme-based simulation model, which is explained in detail below. The parameters and their values used in the simulation of the BFO are given in Table.\ref{tab:Sim1}.

\begin{table}[b]
\centering
\begin{tabular}{|l|c|c|}
\hline
\textbf{Physical quantity} & \textbf{Symbol} & \textbf{Value} \\
\hline
Temperature (K) & $T$ & 298 \\
\hline
Phonon frequency (Hz) & $\nu_{0}$ & $1 \times 10^{12}$\\
\hline
Lattice constant (m) & $d$ & $0.56 \times 10^{-9}$\\
\hline
Device area ($\rm mm^{2}$) & $A_{\rm d}$ & $0.04$\\
\hline
Relative permittivity of $\rm BiFeO_{3}$ & $\varepsilon_{r} $ & 52.0\\
\hline
Conductivity of $\rm BiFeO_{3}$ ($\Omega {\rm m}$) & $\sigma$ & $7.0 \times
 10^{-4}$\\
 \hline
 Length of BFO (m)& $l_{\rm BFO}$ & $600 \times 10^{-9}$\\
 \hline
Defect density (${\rm cm^{-3}}$) & $\rho$ & $2\times 10^{16}  $\\
\hline
Top Schottky barrier height (eV)& $\Phi^{\rm t}_{\rm 0}$ & $0.8$\\
\hline
Top Schottky barrier ideality factor & $n_{0}^{\rm t}$ & 4.2\\
\hline
Bottom Schottky barrier height (eV) & $\Phi^{\rm b}_{\rm 0}$ & $0.85$\\
\hline
Bottom Schottky barrier ideality factor & $n_{0}^{\rm b}$ & 4.5\\
\hline
\end{tabular}
\caption{\label{tab:Sim1}Details of the simulation parameters~\cite{Du2018}}
\end{table}

The simulation approach is adapted from Yarragolla \textit{et al.}~\cite{Yarragolla2022}. The approach combines Newton's laws of motion with the kinetic cloud-in-a-cell (CIC) scheme to couple the vacancy transport with the electric field in the $\rm BiFeO_{3}$ layer. The simulation scenario is depicted in Fig.\ref{fig:Sim1b}. The figure shows a BFO modelled on a 1D computational grid with equally and randomly arranged mobile positively charged and fixed negatively charged defects. The negatively charged defects are the fixed oxygen ions ($\rm V_{O}^{-}$), assumed only for having a neutral charge in BFO. 

Initially, all parameters listed in Table.\ref{tab:Sim1} are mostly constants and are defined manually, except for parameters such as defect density and length of the device, which vary slightly depending on the device fabrication process. These variables are selected from a collection of randomly generated values with a truncated Gaussian distribution. Then an equal number of positively charged and negatively charged defects are randomly distributed across the computational domain. The initial activation energy $(U_{\rm A})$ throughout the computational domain is also set manually so that it gradually changes from 0.55\,eV at the Au interface to 0.75\,eV at the Pt/Ti interface. After the initialisation process is complete, the input voltage ($V_{\rm Device}$) is specified, and the effective activation energy, which changes as a function of $V_{\rm Device}$, is calculated using the following line equation:

\begin{ceqn}\begin{align} 
    U_{\rm A, eff} = U_{\rm A} + \lambda_{\rm U}V_{\rm Device}\left ( 1-\frac{x}{l_{\rm BFO}} \right ), 
    \label{Eq:Sim8}
\end{align} \end{ceqn}

\noindent where $x$ is a position in the BFO layer, $l_{\rm BFO}$ is the length of BFO, and $\lambda_{\rm U}$ is the fitting parameter that determines the rate of change of $U_{\rm A,eff}$. $\lambda_{\rm U}$ can be any number between 0 and 1. In the next step, the electric potential ($\phi$) and electric fields ($E$) within BFO are then calculated using the following equations, 

\begin{ceqn}\begin{align} 
    \frac{d}{dx} \left ( \varepsilon \frac{d\phi}{dx}\right ) = -\rho,
    \label{Eq:Sim1}
\end{align} \end{ceqn}

\begin{ceqn}\begin{align} 
E = -\frac{d\phi}{dx}.
\label{Eq:Sim2}
\end{align} \end{ceqn}

\noindent Here, $\rho$ is the charge density, and $\varepsilon$ is the permittivity of BFO. For this, Dirichlet boundary conditions are used, which are calculated using the voltage drops at the top and bottom interfaces by applying Kirchhoff's voltage law and Kirchhoff's current law to the equivalent circuit shown in Fig.\ref{fig:Sim1a}\cite{Yarragolla2022}.

\begin{ceqn}\begin{align} 
    V_{\rm Device} = V_{\rm SC,t} + V_{\rm BFO} - V_{\rm SC,b}, 
    \label{Eq:Sim3}
\end{align} \end{ceqn}

\begin{ceqn}\begin{align} 
    I_{\rm SC,t} = I_{\rm BFO} = -I_{\rm SC,b} = I,
    \label{Eq:Sim4}
\end{align} \end{ceqn}

 \noindent  where $V_{\rm SC,t/b}, V_{\rm BFO}$ are the voltage drops across the top and bottom Schottky barrier and BFO, respectively. Similarly, $I_{\rm SC, t/b}, I_{\rm BFO}$ are the currents across the top and bottom Schottky barrier and BFO, respectively. An iterative scheme mentioned by Yarragolla et al. is used to calculate the voltage drops and currents. Followed by this, the electric potentials at Au/BFO and BFO/Pt interfaces are given by $\phi_{\rm Au/BFO} = V_{\rm Device} - V_{\rm SC,t}$ and $\phi_{\rm BFO/Pt/Ti} = -V_{\rm SC,b}$, respectively.
 
 The currents mentioned in Eq.\eqref{Eq:Sim4} can be calculated as follows. The resistive switching in BFO is mainly attributed to the change in Schottky barrier properties at the metal/oxide interfaces. As seen in Fig. \ref{fig:Sim1a}, both diodes are initially in rectifying mode with forward-biased $\rm D_{t}$ and reverse-biased $\rm D_{b}$. When a positive biased is applied, the $\rm V_{O}^{+}$ vacancies drift in the presence of an electric field, changing the $\rm V_{O}^{+}$ vacancies concentration at the BFO/Pt interface. This change in $\rm V_{O}^{+}$ vacancy concentration significantly decreases the barrier height of $\rm D_{b}$, thus making it conducting. $\rm D_{b}$ is forward biased during a RESET process and changes its mode from conducting to rectifying, while $\rm D_{t}$ is reversed biased and still in the rectifying mode. 

In general, the current through the Schottky barrier is determined by thermionic emission (TE), thermionic field emission (TFE), direct tunnelling or a combination of these mechanisms. However, for a BFO memristive device with strong rectifying properties due to the two metal-semiconductor contacts at the top and bottom, it is sufficient to consider the mechanisms of thermionic emission and thermionic field emission. Since direct tunnelling does not lead to rectifying properties, it can be excluded as one of the dominant mechanisms. Therefore, the current across $\rm D_{t}$ and $\rm D_{b}$ is calculated using the Richardson-Schottky equation~\cite{Sze2007}, which is based on the TE mechanism and further modified by considering an effective ideality factor $(n_{\rm eff})$ greater than one to account for the additional current tunneling through the barrier (i.e., described by TFE)\cite{Sze2007,Tyagi1984}. The final current equation for the Schottky barriers is given by

\begin{ceqn}\begin{align} 
    I_{\rm SC} = I_{\rm R}\left ( {\rm exp}\left \{ \frac{eV_{\rm SC}}{n_{\rm eff}k_{B}T} \right \} - 1\right ).
    \label{eq:Current1}
\end{align} \end{ceqn}

\noindent The reverse current, $I_{\rm R}$ for forward and reverse bias conditions are respectively given by

\begin{ceqn}\begin{align} 
    I_{\rm R, V_{SC}> 0} = A_{\rm d}A^{*} T^{2}{\rm exp}\left \{ \frac{-\Phi_{\rm eff}}{k_{B}T} \right \}, \hspace{0.1cm}{\rm and}
    \label{eq:Current2}
\end{align} \end{ceqn}

\begin{ceqn}\begin{align} 
        I_{\rm R, V_{SC}<  0} = A_{\rm d}A^{*} T^{2}{\rm exp}\left \{ \frac{-\Phi_{\rm eff}}{k_{B}T} \right \}
        {\rm exp}\left \{ \frac{\alpha _{r}\sqrt{\left | V_{\rm SC} \right |}}{k_{B}T} \right \}.
        \label{eq:Current3}
\end{align} \end{ceqn}

\noindent Here $k_{B}$ is the Boltzmann constant, $T$ is the temperature, and $A^{*} \rm{=  1.20173 \times 10^{6} A/ (m^{2}K^{2})}$ is the effective Richardson constant. The effective ideality factor, $n_{\rm eff} = n_{\rm 0}(1 + \lambda_{n}\,q(t))$ and the effective Schottky barrier height, $\Phi_{\rm eff} = \Phi_{\rm 0}(1 + \lambda_{\rm b}\, q(t))$, depend on the internal state of the device, $q{(t)}$. $\Phi_{\rm 0}$ is the initial Schottky barrier height, and $n_{0}$ is the initial ideality factor. $\lambda_{\rm b}$ and $\lambda_{\rm n}$ are the fitting parameters that define the rate of change of effective barrier height and effective ideality factor, respectively.

           \begin{figure}[t]
                \centering             \includegraphics[width=0.98\textwidth]{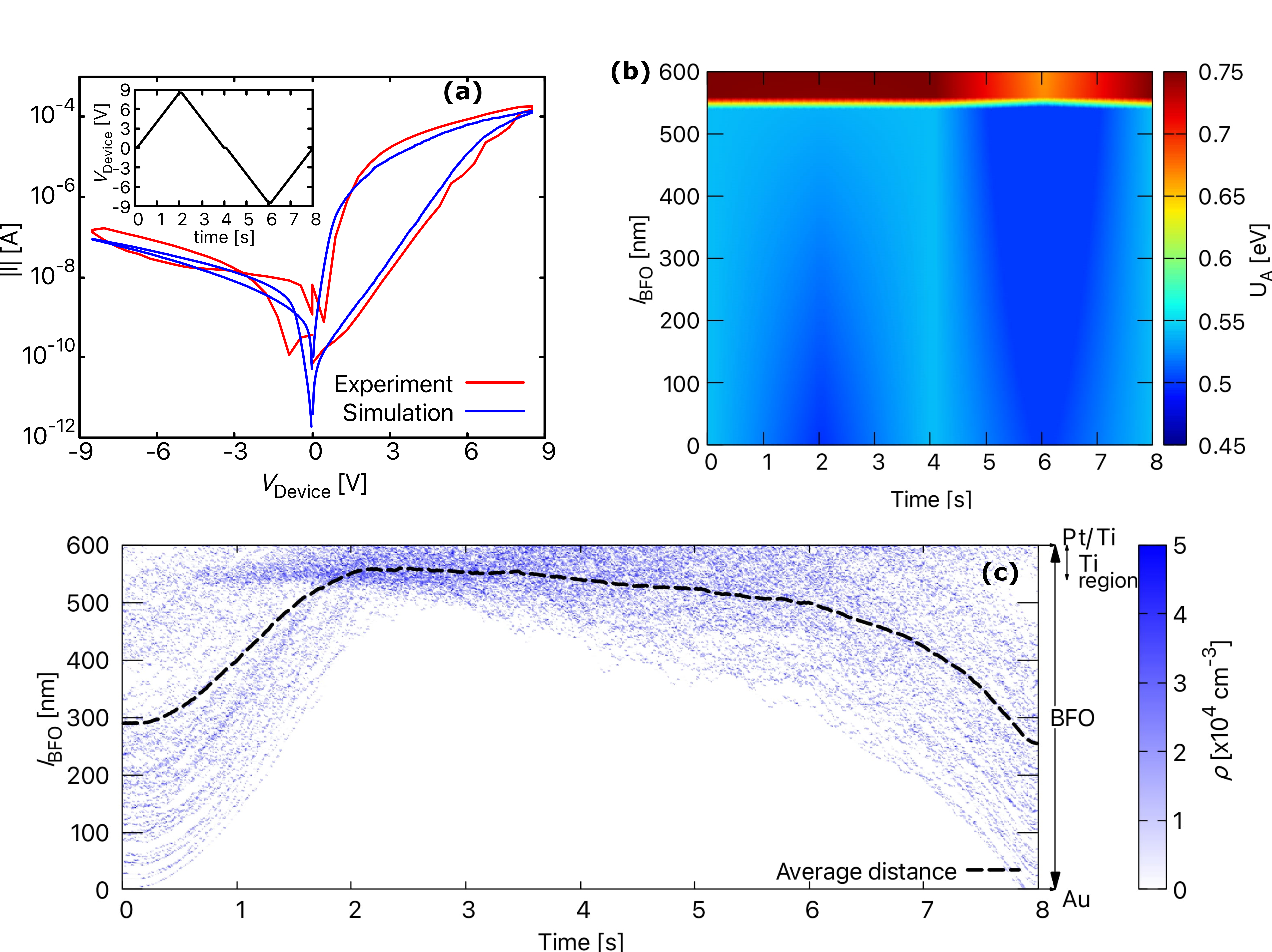}
                \caption{The resistive switching process in BFO memristor. (a) calculated and experimentally obtained \textit{I}-\textit{V} characteristics of the BFO memristor, (b)the change in activation energy with time across the BFO device during the set and reset process, and (c) the ion transport in BFO memristor shown as a change in charge density ($\rho$) over time. The black dashed line indicates the absolute average distance $\bar{d}(t)$}
                \label{fig:results3}
            \end{figure}

The current flowing through the central BFO region can be calculated by applying Ohm's law. 

\begin{ceqn}\begin{align} 
    I_{\rm BFO} = \sigma A_{\rm d}\frac{V_{\rm BFO}}{l_{\rm BFO}},
    \label{eq:Current4}
\end{align} \end{ceqn}

\noindent where $\sigma$ is the conductivity of BFO and $l_{\rm BFO}$ is the length of BFO. Using Eq.\eqref{Eq:Sim4}, the above Eq.~\eqref{eq:Current4} can be modified as

\begin{ceqn}\begin{align} 
    V_{\rm BFO} = I_{\rm SC,t}\frac{l_{\rm BFO}}{\sigma A_{\rm d}}.
    \label{eq:Current5}
\end{align} \end{ceqn}

Using the electric field from Eq.\eqref{Eq:Sim2} and the effective activation energy, the drift velocity of $\rm V_{O}^{+}$  is given as follows\cite{bruce_1994,Meyer2008}:

\begin{ceqn}\begin{align} 
    v_{\rm D} = \nu_{0} d  {\rm exp}\left ( -\frac{{U}_{\rm A,eff}}{k_{\rm B}T} \right )\left ( {\rm exp}\left \{ \frac{\left | z \right |edE}{2k_{\rm B}T}
 \right \} - {\rm exp}\left \{ -\frac{\left | z \right |edE}{2k_{\rm B}T}
 \right \}\right ),
 \label{Eq:Sim5}
\end{align} \end{ceqn}

\noindent where $\nu_{0}$ is the phonon frequency, $U_{\rm A,eff}$ is  the  effective activation  energy, $d$ is the jump distance (lattice constant), $z$ is the charge number, $k_{B}$ is the Boltzmann constant, $T$ is the temperature, and $e$ is the elementary charge. Solving this velocity equation, we obtain a deterministic position of the $\rm V_{O}^{+}$ vacancies. But, in order to account for the intrinsic stochastic behaviour of the BFO device, the position of every $i^{\rm th}$ $\rm V_{O}^{+}$ vacancy is artificially perturbed and given by

\begin{ceqn}
\begin{align}
    \underset{\rm stochastic}{\underbrace{\bar{x}_{i}}} = \underset{\rm deterministic}{\underbrace{\bar{x}_{i}}} +  \hspace{5pt}\underset{\rm stochastic}{\underbrace{(r-0.5)\delta \bar{x}_{i}}}.
    \label{Eq:Sim6}
\end{align}
\end{ceqn}

\noindent Here $r$ is the random number between 0 and 1, and $\delta$ is the  maximum random displacement. After every iteration of the ion movement, the average relative distance between the current position of the vacancies, $\bar{d}(t)$ and their initial position, $\bar{d}_{\rm r}$, determines the internal state of the device,

\begin{ceqn}
\begin{align}
       q(t) = \frac{\bar{d}(t)-\bar{d}_{\rm r}}{\bar{d}_{\rm r}}.
       \label{Eq:Sim7}
\end{align}
\end{ceqn}


\section*{Results and discussion}

           \begin{figure}[!t]
                \centering             \includegraphics[width=0.99\textwidth]{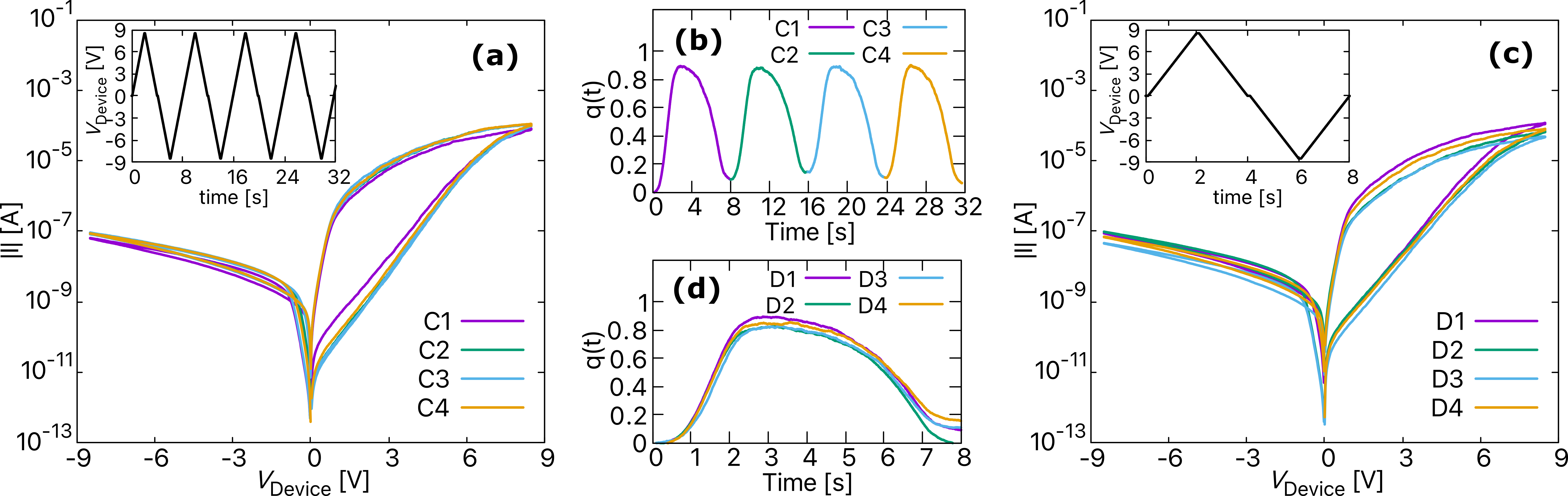}
                \caption{The simulated IV curves showing (a) cycle-to-cycle (C2C) variability obtained for four consecutive voltage sweeps with initial conditions given in the first row of Table.\ref{tab:Sim2}. (b) The change in internal state of the device ($q(t)$) for C2C. (c)~device-to-device (D2D) variability obtained for a single voltage sweep but with different initial conditions given in Table.\ref{tab:Sim2}. (d) The change in internal state of the device ($q(t)$) for D2D. The maximum applied voltage in both plots is $\pm$8.5\,V}
                \label{fig:results4}
            \end{figure}

%

First, the 1D compact model of BFO device described above was validated by comparing the calculated current-voltage characteristics (\textit{I}-\textit{V} curve) with the experimentally obtained \textit{I}-\textit{V} curve. For this purpose, the simulation model was initialized with the parameters listed in Table \ref{tab:Sim1}. In this state, the device is in its HRS. A voltage sweep with a ramp from 0\,V through 8.5\,V to 0\,V was applied to set the device to an LRS, and then from 0\,V through -8.5\,V to 0\,V to reset it back to HRS. For both set and reset process, a sweeping velocity of 0.36\,V per 100\,ms was applied. The change in applied voltage ($V_{\rm device}$) with time and the resultant \textit{I}-\textit{V} curves are shown in Fig.\ref{fig:results3}(a). From the \textit{I}-\textit{V} curves, it can be seen that the model reproduces the analogue behaviour of BFO very well. The calculated \textit{I}-\textit{V} curve for the SET process (0\,V to 8.5\,V to 0\,V) agrees with the experimental results both qualitatively and quantitatively. However, although the \textit{I}-\textit{V} curve for RESET agrees quite well quantitatively with the experimental \textit{I}-\textit{V} curve, the model does not entirely capture the effect of non-zero crossing (at -3\,V) observed in the experimental \textit{I}-\textit{V} curve. Sun \textit{et al.}\cite{Sun2020} attributed this type of non-zero crossing behaviour of memristors to three mechanisms. They are the capacitive effect, the ferroelectric polarisation effect, and the presence of an internal electromotive force. Since the ferroelectric polarisation effects are already ruled out for a BFO, the possible reason for non-zero crossing in a BFO can most likely be the capacitive effects. The capacitance-voltage measurements of Shuai \textit{et al.}~\cite{Shuai2013,Shuai2018} showed the presence of such capacitive effects in BFO. They indicated the presence of simultaneous resistive and capacitive switching in BFO, with HRS corresponding to the low capacitance state and vice versa.

           \begin{figure}[t]
                \centering             \includegraphics[width=0.99\textwidth]{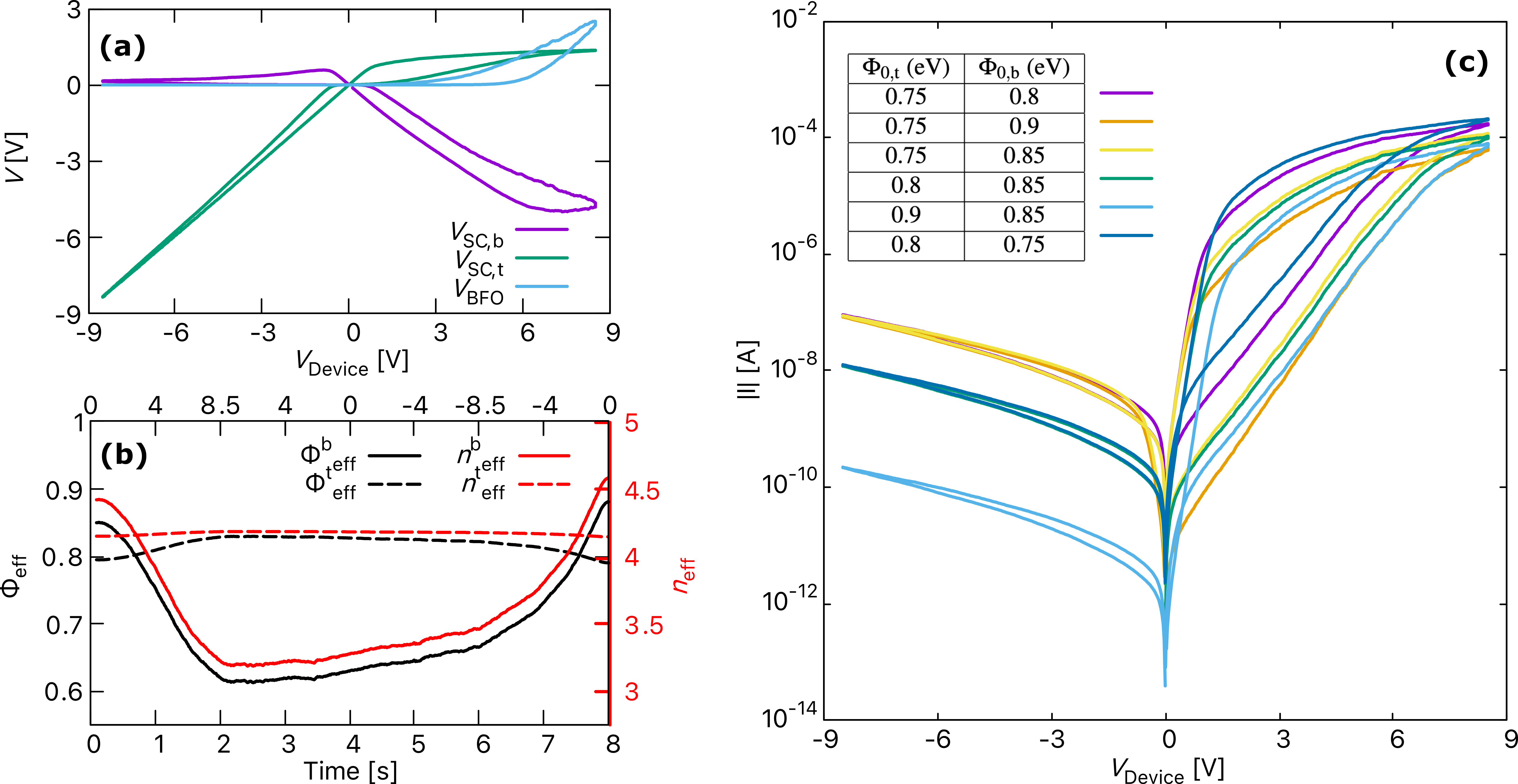}
                \caption{(a) Voltage drop across different regions of the BFO memristor. (b) The change in top and bottom effective Schottky barrier heights ($\Phi_{\rm eff}$) and effective ideality ($n_{\rm eff}$) factor as a function of time. (c) The simulated \textit{I}-\textit{V} curves showing the effect of top and bottom Schottky barrier height on the set and reset process. Curves with $\Phi_{\rm 0,t}$=0.8\,eV overlap in the negative bias region.}
                \label{fig:results5}
            \end{figure}


\begin{figure}[b!]
\begin{minipage}{.45\textwidth}
    \centering 
\begin{tabular}{|c|c|c|c|}
\hline
Device & $\bar{d_{\rm r}}$ (nm)& $l_{\rm BFO}$ (nm) & $\rho$ ($\rm cm^{-3}$) \\
\hline
1 & 295.4 & 600 & $ 2\times 10^{16}$  \\
\hline
2 & 306.89 & 588.2 & $ 2.6 \times 10^{16}$\\
\hline
3 & 299.001 & 601.5 & $ 2.1 \times 10^{16}$\\
\hline
4 & 302.23 & 586.9 & $ 1.5 \times 10^{16}$\\
\hline
\end{tabular}%
\captionof{table}{\label{tab:Sim2} The parameters of the four devices used to determine the \textit{I}-\textit{V} curves in Fig.\ref{fig:results4}}
\end{minipage}\hfill\hspace{0.8cm}
\begin{minipage}{.55\textwidth}
    \centering 
\begin{tabular}{|c|c|c|c|}
\hline
T (K) & $\varepsilon_{r}$ & $\sigma$ ($\Omega {\rm m}$) & $\lambda_{\rm T}$\\
\hline
298 & 52 & $8\times10^{-4}$ & 0.0 \\
\hline
313 & 60 & $9\times10^{-4}$ & 0.005\\
\hline
323 & 72 & $1\times10^{-3}$ & 0.01\\
\hline
333 & 100 & $3.5\times10^{-3}$ & 0.04\\
\hline
343 & 132 & $5\times10^{-3}$ & 0.06\\
\hline
348 & 145 & $7\times10^{-3}$ & 0.062\\
\hline
\end{tabular}%
\captionof{table}{\label{tab:Sim3} The parameters used to simulate the\\
\hspace{-3.8cm}\textit{I}-\textit{V} curves in Fig.\ref{fig:results6}}
\end{minipage}
\end{figure}

          \begin{figure}[!t]
                \centering             \includegraphics[width=0.99\textwidth]{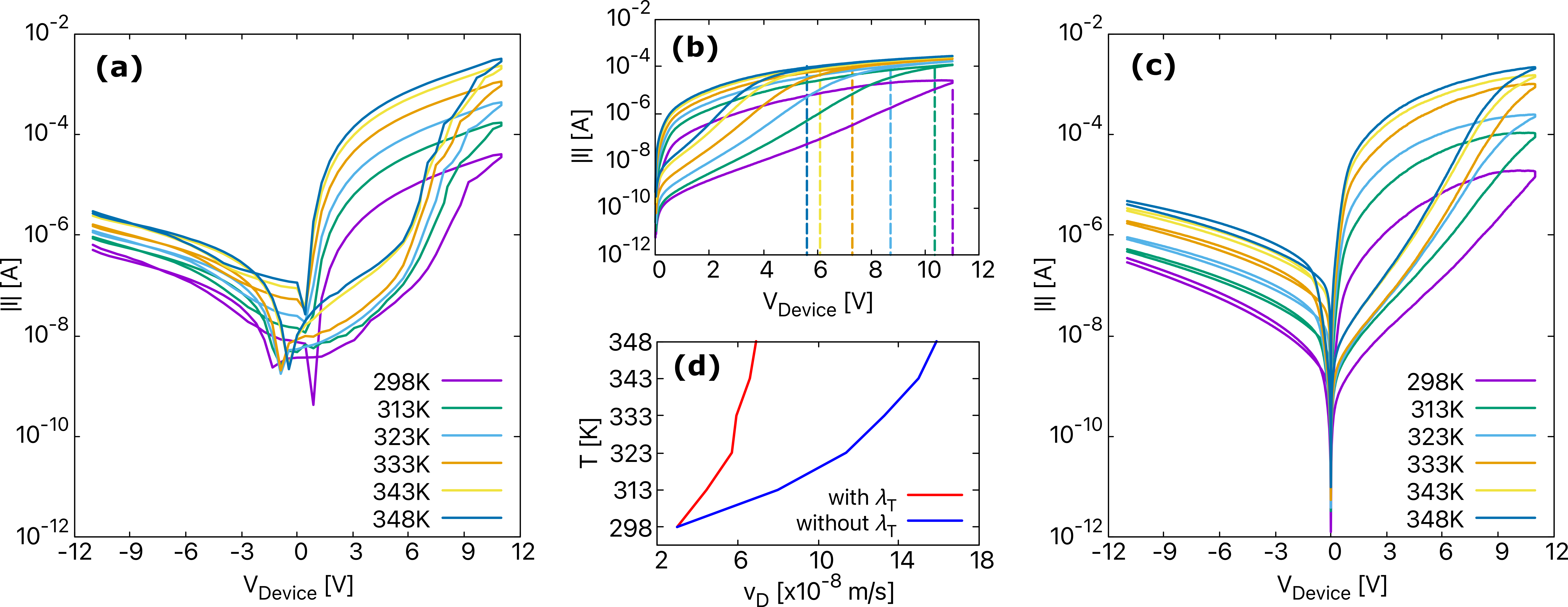}
                \caption{(a) The experimentally obtained temperature-dependent current-voltage characteristics (\textit{I}-\textit{V}-\textit{T} curves). (b) Simulated \textit{I}-\textit{V}-\textit{T} curves obtained using the fitting parameter ($\lambda_{\rm T}$). The parameters used in the simulation are given in Table.\ref{tab:Sim3}. (c) \textit{I}-\textit{V}-\textit{T} curves obtained for positive applied voltage and without $\lambda_{\rm T}$. The dashed lines indicate the voltage required to switch the device from HRS to LRS. The legend is same as mentioned in Fig.\ref{fig:results6}(b). (d) The calculated average drift velocity ($\nu_{\rm D}$) with and without $\lambda_{\rm T}$ at different temperatures.}
                \label{fig:results6}
            \end{figure}

Fig.\ref{fig:results3}(b) shows the change in activation energy ($U_{\rm A}$) across the BFO memristor as a function of time and $V_{\rm device}$ shown in Fig.\ref{fig:results3}(a). As mentioned earlier, the activation energy across the BFO is not homogeneous but gradually increases from 0.55\,eV to 0.75\,eV between 550\,nm and 600\,nm.This inhomogeneous $U_{\rm A}$ is due to the presence of the Ti region between 550\,nm and 600\,nm and plays an essential role in the vacancy transport responsible for the resistive switching behaviour of BFO. Vacancy transport is illustrated in Fig.\ref{fig:results3}(c) by considering how the charge density in BFO changes during the simulation time of 8\,s. Initially, the vacancies are randomly placed across the BFO computational domain, and when a voltage is applied, the vacancies move towards the BFO/Pt interface. Due to the very high activation energy (about 0.75\,eV) at the BFO/Pt interface, the change in position of the $\rm V_{O}^{+}$ vacancies near the Au/BFO interface is insignificant. This insignificant change in the position of the $\rm V_{O}^{+}$ vacancies between 2\,s and 3.2\,s sets the device to an LRS with a nearly constant average relative distance ($\bar{d}(t)\approx562$\,nm). Moreover, the change in the position of the $\rm V_{O}^{+}$ vacancies is so small that they can be assumed to be almost in a trapped state. When a negative write bias is applied after 4\,s during the reset process, the activation energy at the BFO/Pt interface drops to about 0.6\,eV (as shown in Fig.\ref{fig:results3}(b)), which is sufficient to return the $\rm V_{O}^{+}$ vacancies to their equilibrium position and reset the device to HRS. At the end of the reset process (at 8\,s), the average relative distance also drops to an initial value of about 268\,nm.


The intrinsic stochastic behaviour of BFO is measured in terms of spatial (device to
 device) and temporal (cycle to cycle) variability. First, the four \textit{I} -\textit{V} curves in Fig.\ref{fig:results4}(a) show the temporal variability of the BFO. The curves were obtained for four consecutive voltage sweeps shown in the inset of Fig.\ref{fig:results4}(a), and the initial conditions used to simulate these curves are given in the first row of Table.\ref{tab:Sim2}. Although, the \textit{I} -\textit{V} curves calculated for each applied voltage cycle follow a similar trend, they slightly vary from each other. As observed from Fig.\ref{fig:results4}(b), the slight variation in the \textit{I} -\textit{V} curves is likely due to the change in internal state of the device ($q(t)$) from cycle to cycle. This change in $q(t)$ is actually due to the random movement of the vacancies as calculated using Eq.\eqref{Eq:Sim6}. Second, the spatial variability in BFO is illustrated using the four \textit{I} -\textit{V} curves in Fig.\ref{fig:results4}(c) for a single voltage sweep shown in the inset. The initial conditions used to simulate the four curves are given in Table.\ref{tab:Sim2}. As can be observed, the \textit{I} -\textit{V} curves showing spatial variability are more clearly separated from each other than the \textit{I} -\textit{V} curves showing temporal variability. In general, based on experimental observations, the temporal variability, is much lower for interface-type memristive devices such as BFO and DBMD\cite{Yarragolla2022}. Moreover, the difference in the \textit{I} -\textit{V} curves showing spatial variability could be mainly due to the different initial conditions used to simulate each curve. The initial conditions more or less directly effect $q(t)$, so for this reason we observe a change in $q(t)$ as seen in Fig.\ref{fig:results4}(d).  However, apart from the above-mentioned reasons, it is predicted that, several other unknown physical or chemical phenomena may also contribute to this spatial and temporal variability in a true BFO memristor.

The voltage drops across different BFO memristor regions are plotted in Fig.\ref{fig:results5}(a) to investigate the rectification process and the physical mechanisms behind the strong voltage dependence. The different regions include the two Schottky contacts at the Au/BFO ($\rm D_{t}$) and BFO/Pt ($\rm D_{b}$) interfaces, and the $\rm BiFeO_{3}$ layer. For a positive bias, a back-to-back rectification is observed with a forward-biased $\rm D_{t}$ and reverse-biased $\rm D_{b}$. As can be seen in Fig.\ref{fig:results5}(a), although there is some voltage drop across the $\rm D_{t}$ and $\rm BiFeO_{3}$ layer, most of the voltage is blocked by the reverse-biased $\rm D_{b}$. Furthermore, for a negative bias, almost all the applied voltage is blocked by the reverse-biased $\rm D_{t}$, with a small voltage drop across the $\rm D_{b}$ and a negligible one across the BFO layer.

The changes in other properties of the Schottky contacts, such as the barrier height ($\Phi$) and the ideality factor ($n$) during the sweep time, are also shown in Fig.\ref{fig:results5}(b). The change in $\Phi$ and $n$ are strongly influenced by the vacancy transport in the BFO. As vacancies move toward the BFO/Pt interface during positive bias, $\rm D_{b}$ becomes non-rectifying, with $\Phi_{\rm eff}^{\rm b}$ decreasing from 0.85\,eV to 0.62\,eV and $n_{\rm eff}^{\rm b}$ from 4.5 to 3.3. This allows electrons to flow easily across the barrier, increasing the current through the BFO memristor. With a negative bias, $\Phi_{\rm eff}^{\rm b}$ and $n_{\rm eff}^{\rm b}$ increase as the vacancies move away from the BFO/Pt interface and $\rm D_{b}$ becomes rectifying. On the other hand, $\Phi_{\rm eff}^{\rm t}$ and $n_{\rm eff}^{\rm t}$ increase with a positive bias and decrease with a negative bias. However, their change is almost negligible; therefore, $\rm D_{\rm t}$ can be considered non-flexible and constantly rectifying.

Moreover, the initial Shottky barrier heights are considered as one of the entropy sources in hardware security applications. So, it is also important to check the behaviour of a BFO memristor to change in  initial Schottky barrier heights. The graph in Fig.\ref{fig:results5}(c) shows the \textit{I} -\textit{V} curves with different combinations of $\Phi_{0}^{\rm t}$ and $\Phi_{0}^{\rm b}$. Ideally, the barrier height can be increased or decreased by reducing or increasing the doping concentration, respectively\cite{Hudait2001}. Increasing $\Phi_{0}^{\rm t/b}$ from 0.75\,eV to 0.9\,eV increases the energy required for the electrons to cross the barrier, which reduces the current through the BFO. In contrast, if $\Phi_{0}^{\rm t/b}$ is decreased, the electrons can move more easily, which increases the current through the BFO. Also, a change in $\Phi_{0}^{\rm b}$ mainly affects the right loop of the \textit{I} -\textit{V} curves, while a similar change in $\Phi_{0}^{\rm t}$, affects the left loop of the \textit{I} -\textit{V} curve. Therefore, as mentioned by Du \textit{et al.}\cite{Du2018} and observed in Fig.\ref{fig:results5}, we can conclude that the set current is mainly determined by the barrier height and the voltage drop across $\rm D_{b}$, and the reset current by the barrier height and the voltage drop across $\rm D_{t}$.


\begin{figure}[t]
\begin{center}
    \includegraphics[width=1.0\textwidth]{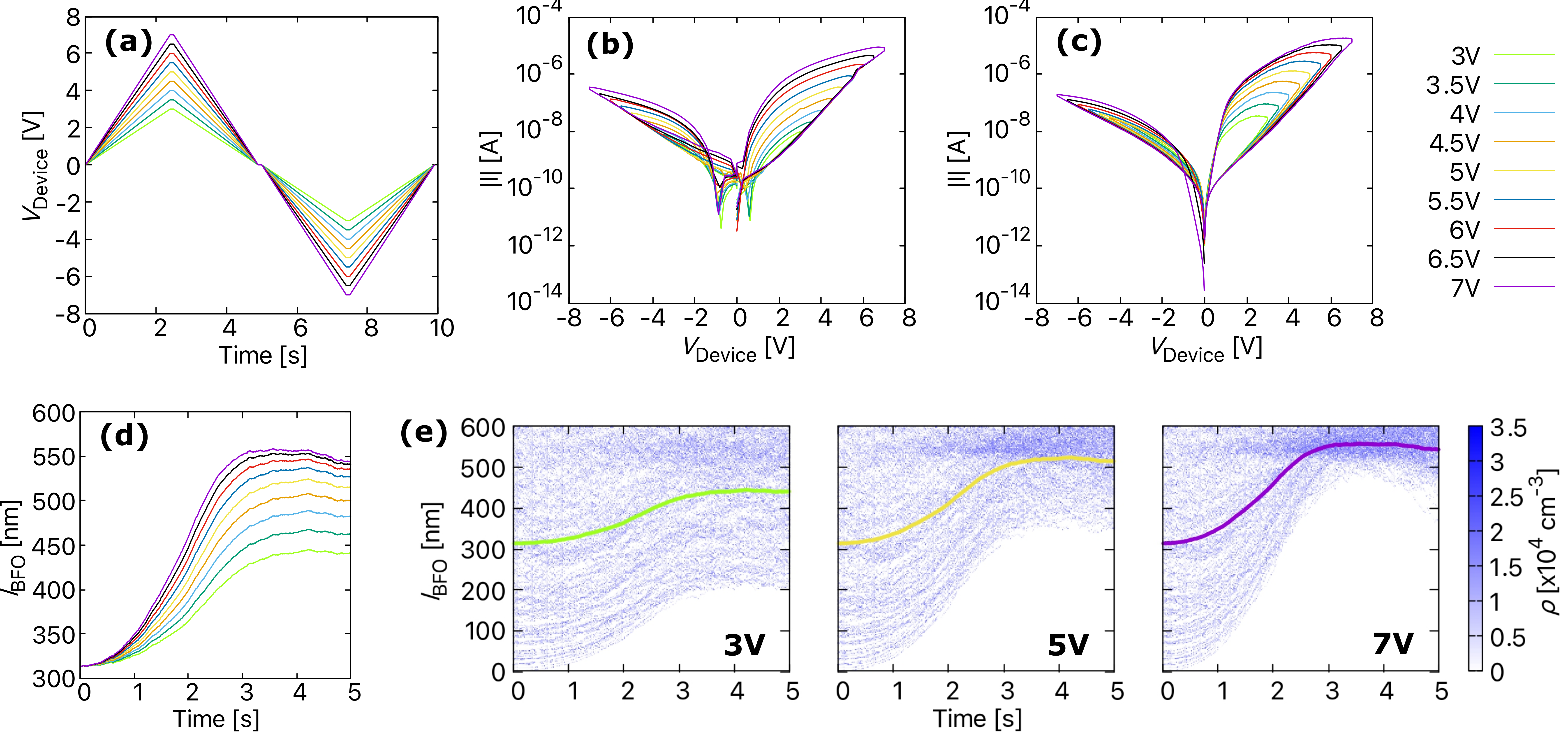}
\caption{(a) Input voltage source ($V_{\rm Device}$) with different amplitudes (b) Experimental \textit{I} -\textit{V} curves, and (c) simulated \textit{I} -\textit{V} curves for different maximum $V_{\rm Device}$. (d) The change in average distance $\bar{d}(t)$ with time for a positive applied voltage cycle with different amplitudes. (e)The change in charge density over time for a positive applied voltage cycle with maximum voltage of 3\,V, 5\,V and 7\,V. The legend for all plots is shown on the top right corner.}
\label{fig:results7} 
\end{center}
\end{figure}

The response of BFO memristor to changing environmental conditions (e.g., temperature) and stress (e.g., excessive voltage) is illustrated in Figs.\ref{fig:results6} and \ref{fig:results7}. These factors are important when considering BFO for neuromorphic circuits, hardware security and non-volatile memory applications. Fig.\ref{fig:results6} shows the simulated and experimental temperature-dependent $ \textit{I} $-$ \textit{V} $ curves. The results are obtained for a writing bias of $\pm$11\,V and different temperatures increasing from 298\,K to 348\,K. According to Eq.\eqref{Eq:Sim5}, the velocity of the vacancies ($\nu_{\rm D}$) increases with increasing temperature. As $\nu_{\rm D}$ increases, the maximum voltage required to switch the device to LRS reduces as shown in Fig.\ref{fig:results6}(b). However, as observed experimentally in Fig.\ref{fig:results6}(a), the maximum voltage required for switching is the same for all temperatures. No change in the switching voltage suggests that there might be some frictional force acting on the vacancies that decrease their velocity ($\nu_{\rm D}$) with increasing temperature. This frictional force could be due to the following reasons: (a) With the increasing temperature, more $\rm Ti^{4+}$ ions may diffuse into the BiFe$\rm O_{3}$ layer due to thermal diffusion. The more Ti diffuses into the BFO layer, the higher the activation energy, resulting in a decrease of $\nu_{\rm D}$, (b) The collision rate between particles increases with increasing temperature, which affects the motion of particles in a device, and (c) the presence of temperature-dependent ferroelectric polarisation switching. To account for the frictional force in BFO, we modify $\nu_{\rm D}$ by using a fitting parameter to match the experimental results. So, $\nu_{\rm D}$ = $\nu_{\rm D}$(1-$\lambda_{\rm T}$) where $\lambda_{\rm T}$ could be any number between 0 and 1. In this way, we can reduce $\nu_{\rm D}$ of all vacancies for different temperatures. This can be seen in Fig.\ref{fig:results6}(d), which shows the average $\nu_{\rm D}$ with and without including the fitting parameter for different temperatures. The parameters used to simulate  \textit{I} -\textit{V} curves in Fig.\ref{fig:results6}(c) are given in Table.\ref{tab:Sim3}. Therefore, by considering the fitting term, we are able to mimic the temperature-dependent resistive switching in BFO, as shown in Fig.\ref{fig:results6}(c).

The measured and simulated \textit{I} -\textit{V} curves at room temperature for different maximum applied voltages are shown in Fig.\ref{fig:results7}. The voltage profiles used to obtain the plots are shown in Fig.\ref{fig:results7}(a). The observations from Fig.\ref{fig:results7}(b) indicate that the shape of the \textit{I} -\textit{V} curve in the set and reset direction strongly depend on the maximum applied voltage. At higher applied maximum voltages, the shape of the hysteresis is relatively wider compared to the hysteresis obtained at lower maximum $V_{\rm device}$. The simulated \textit{I} -\textit{V} curves in Fig.\ref{fig:results7}(c) also show a broadening of \textit{I} -\textit{V} curves with increasing maximum $V_{\rm device}$ and quantitatively match quite well with the experimental results. However, there is a discrepancy in the simulated \textit{I} -\textit{V} curves between voltages -1.5\,V to 1.5\,V. As mentioned earlier, this discrepancy could be due to capacitive effects or the presence of an internal electromotive force. 

Furthermore, when the applied voltage is very low, the vacancies do not receive enough energy to drift to the BFO/Pt interface, resulting in switching failure, i.e. the device does not switch to LRS and remains in a HRS. This can be seen in Fig.\ref{fig:results7}(d), which shows the change in average distance for different maximum applied voltages. The average distance is only tracked for the positive $V_{\rm-device}$ cycle since we are interested in observing the vacancies drift towards the BFO/Pt interface, which mainly contributes toward switching the device to LRS. For low maximum $V_{\rm device}$ from 3\,V to 5\,V, $\bar{d}(t)$ is less than 520\,nm, which means that certain vacancies do not reach the BFO/Pt interface and are not trapped. This can be better understood from Fig.\ref{fig:results7}(e) that shows the vacancy transport in terms of charge density for a positive applied voltage cycle. The retention time of the device is also severely affected due to the untrapped vacancies in the 3\,V and 5\,V plots.

\begin{figure}[t]
\begin{center}
    \includegraphics[width=0.5\textwidth]{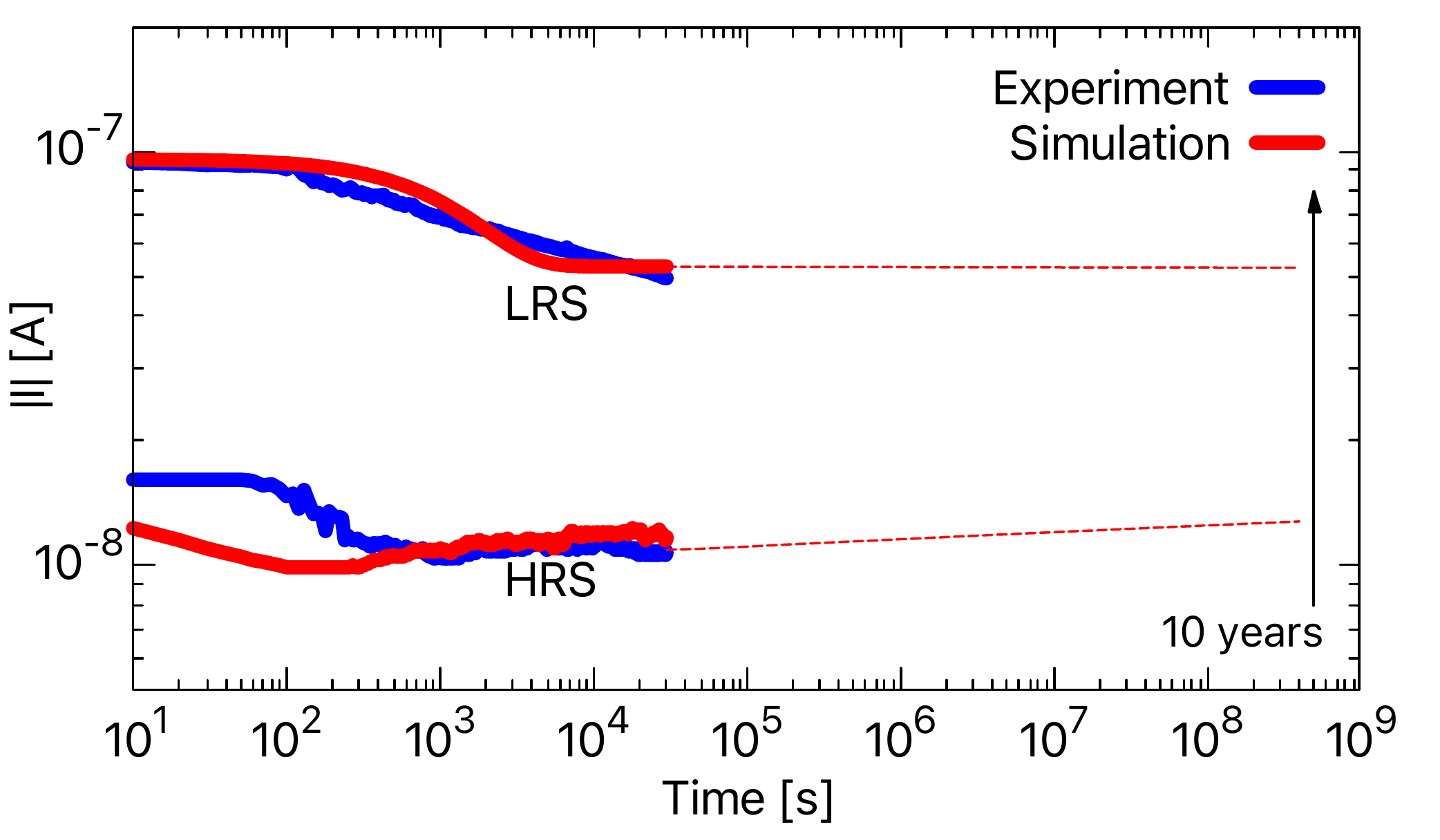}
\caption{The comparison between simulated and experimental retention characteristics of a BFO memristor. The current evolution is recorded every 10\,s for 3000 cycles for both LRS and HRS.}
\label{fig:results8} 
\end{center}
\end{figure}

The retention characteristics of BFO memristor are investigated using the proposed stochastic model. The retention tests were performed by switching the device to LRS or HRS, which are the initial states for this particular study. Then the externally applied voltage was switched off, and the diffusion of the vacancies was recorded. Since the experimental results are obtained for a real BFO device (i.e., 3D), we had to interpolate the 1D model to 3D. In a 3D model, the vacancies can generally move in six directions, while in a 1D space, the vacancies can move either back or forth. So, to fit the simulated results with the experimental results, we use a fitting parameter that restricts the movement of vacancies in BFO by reducing their jump attempts. For this, we randomly pick a number, $\beta$ between 0 and 1, and use the following relation for moving the vacancies:

\begin{equation*}
    \hspace{5cm}\nu_{\rm D}= \nu_{\rm D,\ updated} \ \ \ {\rm for} \  0>\beta<0.33,  
\end{equation*}

\begin{equation*}
\hspace{5cm}\nu_{\rm D} = \nu_{\rm D,\ present}\ \ \  \ {\rm for} \ 0.33>\beta<1.
\end{equation*}

\noindent where $\nu_{\rm D}$ is the drift velocity of the vacancies given by Eq.\eqref{Eq:Sim5}. The development of the device current was recorded every 10\,s with a read voltage of 2\,V at room temperature. The simulated and experimental results for a total of 3000 cycles/pulses are shown in Fig.\ref{fig:results8}. As observed, BFO shows good retention characteristics. The HRS is stable, and no significant change was observed during the 3000 cycles. For the LRS, the good retention is primarily due to the diffusion of $\rm Ti^{4+}$ ions that increases the activation energy at the BFO/Pt interface. This high activation energy limits the movement of vacancies .ie., the vacancies get trapped. However, BFO shows a degradation in the LRS current until approximately the $700^{\rm th}$ cycle, i.e., two hours before stabilizing. Through simulations, this degradation was found to be due to the diffusion of some vacancies away from the BFO/Pt interface that were not trapped. The relative average distance, $\bar{d}(t)$, decreased from 564\,nm to 542\,nm, which increased $\Phi_{\rm eff}^{\rm b}$ from 0.62\,eV to 0.68\,eV, thereby decreasing the current. One possible way to improve retention could be to ensure that all the vacancies are properly trapped by increasing the Ti fluence\cite{You2015}. The second possibility would be to improve the BFO surface using low-energy $\rm Ar^{+}$ ion irradiation, as suggested by Shuai \textit{et al.}\cite{Shuai2013}.

\section*{Conclusion}
With billions of electronic devices connected to the Internet worldwide, low-power, nanoscale memristive devices are considered favourable devices for a more secure Internet of Things. Due to their stochastic behaviour, these memristive devices are ideally suited for hardware security applications such as PUFs, TRNGs, and cryptographic algorithms. The $\rm BiFeO_{3}$-based memristive devices are deemed to be suitable for such applications. In the proposed work, a physics-inspired compact 1D model of an Au/BiFe$\rm O_{3}$ (BFO)/Pt/Ti memristor is developed for circuit-level simulations in the field of hardware security applications and neuromorphic circuits. The model successfully simulates resistive switching based on electric field-driven migration of oxygen vacancies and accounts for the intrinsic stochastic nature of the BFO memristor. A cloud-in-a-cell scheme is used in which Newton's laws are consistently coupled with the Poisson solver. The simulated current-voltage characteristics of the BFO memristor obtained with this scheme agree well with the experimental results. It was found that the set current is mainly determined by the Schottky barrier height and the voltage drop across the BFO/Pt interface, while the reset current is determined by the Schottky barrier height and the voltage drop across the Au/BFO interface. In addition, based on the observations of the simulated and experimental temperature-dependent current-voltage characteristics, we anticipate the presence of a frictional force acting on the oxygen vacancies that increases with temperature. The simulated and experimental results illustrating the effects of temperature, stress, and the retention characteristics of BFO show reasonable agreement. The proposed model is highly efficient and reliable as it consists of various parameters that can be easily tuned to match the experimental results, and the degree of stochasticity can also be adjusted. To further comprehend the switching process in the BFO memristor, the 1D model could be extended to a 2D or 3D model that better represents the real-world BFO memristor.

\section*{Methods}

\textbf{Experiments.} Polycrystalline, 600\,nm thick BFO functional thin film have been grown by pulsed laser deposition on Pt/Ti/SiO$_{2}$/Si substrates. Circular Au top contacts with thickness of 180\,nm have been magnetron sputtered on the BFO thin films using a shadow mask\cite{Schmidt2016,Jin2015}. All the electrical measurements, illustrated in this work, were recorded using a Keithley source meter 2400, which were connected to the PC via GPIB cables and can be controlled through LabVIEW program.\\ 

\noindent\textbf{Simulations.} An in-house model developed using C programming language.

\section*{Data availability}
The datasets generated during and/or analysed during the current study are available from the corresponding author on reasonable request.

\bibliography{sample}

\section*{Acknowledgements}

Y.S., T.H. and T.M.acknowledge the Deutsche Forschungsgemeinschaft (DFG, German Research Foundation) in the frame of Research Grant MU 2332/10-1 and SFB 1461 (Project-ID 434434223). N.D., X.Z., Z.C. and I.P. acknowledge the German Research Foundation (DFG) Projects MemCrypto (Grant Nr. DU 1896/2-1). N.D. acknowledges the Young Scientist Project in SPP 2253 NanoSecurity (DFG PO 1220-18-1). We acknowledge Dr. Danilo Bürger and Prof. Dr. Heidemarie Schmidt for preparing and providing the physical BFO memristors.

\section*{ORCID iDs}
\noindent
S. Yarragolla: \url{https://orcid.org/0000-0002-2973-4943}\\
T. Hemke: \url{https://orcid.org/0000-0003-2436-5840}\\
N. Du: \url{https://orcid.org/0000-0002-7775-7795}\\
T. Mussenbrock: \url{https://orcid.org/0000-0001-6445-4990}


\section*{Author contributions statement}
S.Y. prepared the data sets, implemented the methodology, conducted the simulations, analysed the results, and prepared the first draft of the manuscript. T.M., N.D. and I.P. conceived and directed the conceptual ideas of the work. S. Y., T.H. and T.M. developed the methodological concept for memristor modelling. N.D. designed the experimental work. X.Z. and Z.C. carried out the electrical experiments. T.M. acquired the funding, and administered the project. All authors contributed with interpreting and discussing the results as well as manuscript writing and revision. 


\section*{Competing interests}
The authors declare no competing interests.


\end{document}